 \documentclass[preprint2]{aastex}

\shorttitle{3-Dimensional Core-Collapse}
\shortauthors{Fryer \& Warren}

\begin{document}

\title{Modeling Core-Collapse Supernovae in 3-Dimensions}


\author{Chris L. Fryer\altaffilmark{1} and Michael S. Warren}
\affil{Theoretical Astrophysics, Los Alamos National Laboratory, 
Los Alamos, NM 87545}
\altaffiltext{1}{Feynman Fellow}

\begin{abstract}

We present the first complete 3-dimensional simulations of the 
core-collapse of a massive star from the onset of collapse to 
the resultant supernova explosion.  We compare the structure of 
the convective instabilities that occur in 3-dimensional models 
with those of past 2-dimensional simulations.  Although the 
convective instabilities are clearly 3-dimensional in nature, 
we find that both the size-scale of the flows and the net 
enhancement to neutrino heating does not differ greatly 
between 2- and 3-dimensional models.  The explosion energy, 
explosion timescale, and remnant mass does not differ by 
more than 10\% between 2- and 3-dimensional simulations.

\end{abstract}


\section{Introduction}

Convective instabilities have been invoked to help drive core-collapse 
supernova explosions since Epstein (1979) first argued that negative 
lepton gradients would drive Ledoux convection in the core.  Epstein 
(1979) argued that this convection would increase the transport of 
energy out of the core and help facilitate a supernova explosion.  
Bruenn, Buchler, \& Livio (1979) confirmed that this convection could 
indeed increase the neutrino luminosity and help drive a supernova 
explosion.  Considerable work studying convective instabilities, 
including multi-dimensional models (e.g. Buchler, Livio, \& 
Colgate 1980), followed soon after.  Although entropy gradients 
caused by the shock were suggested during this time (see 
Bruenn, Buchler, \& Livio 1979), Burrows (1987) first suggested 
that this entropy-driven convection could also boost the neutrino 
luminosity and help drive a supernova explosion.

The work of the past two decades has led to two separate convective
regions: one within the extremely dense proto-neutron star core (see
Keil, Janka, \& M\"uller 1996 for a review) and the other in the
region between the proto-neutron star and the accretion shock where
the bounce stalled.  In this latter region, neutrino heating powers an
unstable entropy gradient that drives convection (see Bethe 1990 for a
review).  In the dense proto-neutron star, convection driven by lepton
gradients (Epstein 1979, Keil et al. 1996), entropy gradients (Burrows
1987, Burrows \& Lattimer 1988), and doubly diffusive
(``salt-finger'') instabilities (Mayle \& Wilson 1988) have all been
invoked to increase the neutrino luminosity and hence, the neutrino
heating.  In the neutrino heating region, entropy-driven convection
helps to convert thermal energy gained from neutrino heating into
kinetic energy, improving the over-all efficiency at which neutrinos
from the core deposit energy into the outer layers of the star.  This
latter convection has been studied in a number of 2-dimensional
simulations over the last decade \citep{Mil93,Heretal94,Bur95,Jan96,
Mez98}.

This entropy driven convection occurs shortly after the collapse of
the massive star.  When this core reaches nuclear densities and
nuclear forces rapidly raise the pressure, its collapse halts, sending
a bounce shock through the star.  This bounce shock stalls and leaves
behind an unstable entropy profile that seeds convection in the region
between the proto-neutron star and the edge of the stalled supernova
shock.  Neutrinos leak out from the proto-neutron star and heat this
region, continuing to drive this entropy-driven convection.  It is
this convection that many groups now agree helps drive the supernova
explosion \citep{Heretal94,Bur95,Jan96}.

However, due to limitations in computer hardware and simulation 
software, all of the past work was limited to 2-dimensional
simulations, leaving behind a number of unanswered questions.  Whether
or not this increased efficiency is sufficient to drive a supernova
explosion with the current supernova mechanisms is still a matter of
debate: compare the explosions of \citet{Heretal94} and \citet{Bur95}
to the fizzles of \citet{Mez98}.  A key uncertainty in all of these
simulations lies in the fact that the 2-dimensional simulations are
being used to study an inherently 3-dimensional event in nature.  Some
scientists have argued that nature will produce convective
instabilities that are much different than what we see in the current
2-dimensional simulations.  In other convective problems (e.g. novae) 
it has been found that 2-dimensional models of these inherently 
3-dimensional processes can lead to vastly incorrect answers 
(compare the differences between the 2- and 3-dimensional work 
of Kercek, Hillebrandt, \& Truran 1998, 1999). 

In this letter, we present the first complete 3-dimensional simulations 
of the evolution of a massive star from collapse to explosion, with 
particular emphasis on the differences between 2 and 3-dimensional 
models of the entropy-driven convection.  We follow these simulations 
until a strong supernova shock has been launched, and can hence see 
how these differences affect the final explosion energy, remnant mass, 
and nucleosynthetic yield of these supernovae.

\section{3-Dimensional Simulations}

Our collapse simulations use the smooth particle hydrodynamics (SPH)
technique with the parallel tree algorithm developed by
\citet{War93,War95}.  To this parallel code, we have added the
equation of state and neutrino physics from the supernova code
developed by \citet{Heretal94}.  The equation of state uses the
nuclear equation of state by \citet{Lat91} at high densities and the
\citet{Bli96} equation of state at low densities.  Nuclear burning is
approximated by a nuclear statistical equilibrium scheme
\citep{Hix96}.  The neutrino transport is mediated by the single
energy flux-limiter developed by \citet{Heretal94} with appropriate
geometrical factors for 3-dimensional models.  Details and tests of
this code are described in \citet{War02}.  To facilitate comparison
with past 2-dimensional simulations \citep{Fry99}, the gravity is
calculated assuming a spherically symmetric potential.

Our progenitor is the standard 15\,M$_\odot$ star (s15s7b2) produced by
\citet{Woo95}.  By using the same equation of state for low densities
used by \citet{Woo95}, we can seamlessly map these 1-dimensional progenitors
into our 3-dimensional collapse code.  To study the convection in
detail, we have run 3 core-collapse simulations this progenitor with a
range of resolutions from 300,000 to 3 million particles (see Table
1).  We compare these simulations to past 2-dimensional simulations
which have the same physics implementations \citep{Fry99} to determine
the differences between 2-dimensional and 3-dimensional models of
convection and more fully understand the role convection plays in the
supernova mechanism.

Figure 1 shows the results of models A, B, and C 75\,ms after bounce.
The isosurface shows material with radial velocities of
1000\,km\,s$^{-1}$ and outlines the outward moving convective bubbles.
Between this surface lies the convective downflows.  Note that even in
the high resolution runs, the total number of bubbles is low, roughly
consistent with the number of modes one might expect from the
2-dimensional simulations.  A 2-dimensional slice of model B 
(Fig. 2) reveals convective overturns which are very similar to the
2-dimensional simulations (see Fryer 1999).

The 3-dimensional simulations produce nearly the same neutrino fluxes
and energies that were found in the 2-dimensional simulations
(Fig. 3).  Although this is not surprising because the transport
methods were identical (except for geometrical factors), it does show
that the small differences in the convective motions do not seem to
dramatically affect the neutrino emission.

Given that the 3-dimensional simulations appear similar to the
2-dimensional simulations, it is not surprising that most of the
quantitative results between these simulations are the same.  The
ultimate explosion energy, explosion times, and remnant masses are all
within 10\% of each other.  Although on the surface, the amount of
neutron rich ejecta is similar, the 3-dimensional simulations produce
more extremely low ($Y_e<0.45$) ejecta than the 2-dimensional
models, and 3-dimensional models, if anything, exacerbate the
problem of ejecting too much neutron rich material.

\section{Implications}

Although the structure of the entropy-driven convection in
core-collapse supernova is definitely 3-dimensional, there is close
resemblance between our 3-dimensional simulations and past
2-dimensional simulations.  This suggests that, for the accuracy
currently needed in supernova simulations, 2-dimensional models may be
sufficient to determine the convective enhancement to the
neutrino-driven supernova mechanism.  Certainly, the uncertainties in
the nuclear equation of state and in the neutrino cross-sections and
transport are much larger than the uncertainties caused by assuming
2-dimensional convection.  The fact that the 3-dimensional models
continue to produce too much neutron rich ejecta implies that there
still persists missing pieces to the supernova puzzle (probably the
neutrino transport, neutrino cross-sections, and equation of state are
all culprits).

Our simulations are designed to study the nature of the convection
above the proto-neutron star in core-collapse supernovae.  The
convection arising in our 3-dimensional simulations shows a remarkable
resemblance to 2-dimensional core-collapse simulations.  Unlike the
nova simulations of Kercek, Hillebrandt, \& Truran (1998, 1999), the
structure, extent and energetics of the convection in core-collapse
simulations are the same in 2 and 3-dimensions.
 
We cannot stress enough the fact that the numbers provided in this
paper are not final answers to the collapse of a 15\,M$_\odot$ star.
With better neutrino transport techniques, neutrino cross-sections and
equations of state, these values will change.  However, changes in the
neutrino physics and equation of state are unlikely to change the
nature of the convection above the proto-neutron star.  Unless the
nature of the convection changes dramatically with these improvements,
our 3-dimensional models show that the convection above the
proto-neutron star in core-collapse supernovae is modeled accurately
in 2-dimensions.  Indeed, for the level of convection arising from our
simulations of the core-collapse of a 15\,M$_\odot$ star, we find that
2-dimensional models are sufficiently accurate to model the supernova
mechanism.  Of course, 3-dimensional models will still be essential
for studies of inherently 3-dimensional aspects of core-collapse
(neutron star kicks, gravitational wave emission, etc.).

Note also, that for these simulations (where the gravity is set to be
spherically symmetric), there are no large asymmetries in the
explosion.  Because there are so many convective modes, it is unlikely
that any large asymmetry will develop without some large-scale force
driving that asymmetry (e.g. asymmetries in the neutrino emission or
initial collapse conditions).  It is possible that the convection will
reduce to fewer modes for those explosions with long delays (Janka -
pvt. communication).  If so, it may be possible to produce the
observed neutron star kicks.  But with the current models, convection
alone can not explain the large neutron star kicks.  In future work,
we will address these asymmetry issues by modeling with full gravity
and considering changes to the initial conditions from rotation to
initial asymmetries.

\acknowledgements

We are grateful to H.-T. Janka, A. Hungerford, A. Burrows, S. Woosley,
and W. Miller for useful discussions and support.  This work was
funded by a Feynman Fellowship at LANL, DOE SciDAC grant number
DE-FC02-01ER41176 , and LANL-based ASCI grant.

\begin{deluxetable}{lccccc}
\tabletypesize{\scriptsize}
\tablecaption{Core-Collapse Simulations \label{tbl-1}}
\tablewidth{0pt}
\tablehead{
\colhead{Model} & \colhead{Particle Number} & \colhead{Explosion Energy}\tablenotemark{a}   
& \colhead{T$_{\rm Explosion}$\tablenotemark{b}} & \colhead{M$_{\rm core}$}  &
\colhead{Low $Y_e$ Ejecta\tablenotemark{c}} \\
\colhead{} & \colhead{} & \colhead{($10^{51}$ ergs)}   
& \colhead{s} & \colhead{($M_\odot$)}  &
\colhead{($M_\odot$)}}
\startdata

2D Model\tablenotemark{d} & 15,000 & 3.0 & 0.10 & 1.1 & 0.24 \\
Model A & 300,000 & 2.9 & 0.15 & 1.15 & 0.29 \\
Model B & 1 million & 2.75 & 0.17 & 1.15 & 0.28 \\
Model C & 3 million & $\sim$3\tablenotemark{e} & 0.15 & 1.15 & 0.26 \\

\enddata

\tablenotetext{a}{The explosion energy is as defined in Fryer (1999) 
and includes both internal and kinetic energy.}
\tablenotetext{b}{The explosion time is determined by the duration
after bounce that it takes the shock to reach 1000\,km.}
\tablenotetext{c}{Low $Y_e$ material refers to ejecta with electron
fraction ($Y_e$) beolow 0.49.}  
\tablenotetext{d}{Newtonian gravity simulation from \citet{Fry99}.}
\tablenotetext{e}{Due to the high resolution in Model C, we have 
not yet run the simulation out long enough to accurately determine 
the explosion energy.}

\end{deluxetable}

\end{document}